 \documentstyle[amssymb,aps,prb,multicol,epsfig]{revtex}
\ifpreprintsty
\def\multb{ }
\def\multe{ }
\else
\def\multb{ \begin{multicols}{2}}
\def\multe{ \end{multicols}}
 \fi

\begin{document}
\author{{\bf O. V. Dolgov}$^{1}$ and {\bf M.} {\bf L.} {\bf Kuli\'{c}}$^{2}${\bf \ }}
\address{$^{1}$Max-Planck-Institut, FKF, Heisenbergstr 1, 70569 Stuttgart, Germany\\
$^{2}$Institut f\"{u}r Physik, Theoretische Physik II, Universit\"{a}t\\
Augsburg, \ 86135 Augsburg, Germany}
\title{{\bf Optical Properties of Heavy Fermion Systems with SDW Order}}
\maketitle

\begin{abstract}
The dynamical conductivity $\sigma (\omega )$, reflectivity $R(\omega )$,
and tunneling density of states $N(\omega )$ of strongly correlated systems
( like heavy fermions) with a spin-density wave (SDW) magnetic order are
studied as a function of impurity scattering rate and temperature. The
theory is generalized to include strong coupling effects in the SDW order.
The results are discussed in the light of optical experiments on
heavy-fermion SDW materials. With some modifications the proposed theory is
applicable also to heavy fermions with localized antiferromagnetic (LAF)
order.
\end{abstract}

\multb

\section{\ Introduction}

It is well known, that below a coherence temperature $T_{coh}(\approx 20-100$
$K)$ heavy-fermion (HF) metals, which contain magnetic ions with 4f or 5f
electrons, are characterized by the large quasiparticle mass $m_{hf}$. The
latter is several hundred times larger than the electron mass $m_{e}$ (see,
Ref. \cite{Heavy}). At temperatures below $T_{coh}$ these systems behave
like strongly renormalized Fermi liquids. As a consequence, various
thermodynamic and transport properties are significantly renormalized below $%
T_{coh}$. For instance, there is a significant increase: (1) in the low
temperature specific heat ($C_{p}=\gamma _{hf}T$), where $\gamma _{hf}\sim
10^{3}\gamma $ with typical value for alkali metals$\ \gamma \sim
1mJmol^{-1}K^{-2}$; (2) in the spin paramagnetic susceptibility; (3) in the
electronic Gr\"{u}neisen parameter $\Omega _{hf}=-d\ln T_{coh}/du$ ($u$ is
the strain field), where for instance $\Omega _{hf}$ of the HF metal $UPt_{3}
$ is 100 times larger than in standard alkali metals; (4) in the
thermoelectric power $Q_{hf}\gtrsim 10^{2}$ $Q$ where $Q$ is the value for
the standard alkali metals. The main reason for this renormalization is the
presence of the Abrikosov-Suhl-Kondo (ASK) resonance in the quasiparticle
density of states at the Fermi level (for more details, see \cite{Heavy}). 

The magnetic properties of heavy fermion materials result from the
competition between the Kondo effect - which gives rise to the screening of
the $4f$ $(5f)$ magnetic moment below the coherence temperature $T_{coh}$,
and the RKKY interaction - which usually gives rise to an antiferromagnetic
(AF) ground state \cite{Heavy}. The itinerant SDW  magnetic order is most
probably realized in $UPt_{3}$, $URu_{2}Si_{2}$, $UCu_{5}$, while in $%
UPd_{2}Al_{3}$ it seems that an AF localized magnetic order is present. In
that respect resistivity measurements on $URu_{2}Si_{2}$ show significant
changes in $\rho (T)$ below $T_{N}(\approx 17K)$ pointing to the SDW type of
magnetism, while in $UPd_{2}Al_{3}$ there is only a small kink in $\rho (T)$
below $T_{N}(\approx 14K)$. The latter fact is more in favor of a localized
magnetic order \cite{Degiorgi,Palstra,Geibel}. Note, that neutron scattering
measurements \cite{Sato,Krimmel,Paolasini}, show that the magnetic moment ($%
\mu \sim 0.85$ $\mu _{B}$) in $UPd_{2}Al_{3}$ is much larger than in $UPt_{3}
$, $URu_{2}Si_{2}$, where one has $\mu \sim 0.01$ $\mu _{B}$. The small
magnetic moment is compatible with the SDW order. We stress that at present
there is no microscopic theory of the itinerant SDW or localized magnetic
order in these systems.

Optical measurements in these compounds play a very important role to
clarify physical nature, which gave possibility for the study of progressive
development of the optical gap in spectra of the SDW systems $UPt_{3}$, $%
URu_{2}Si_{2}$- see \cite{Degiorgi,Donovan,Bonn,Degiorgi2}. Infrared optical
measurements in $UPd_{2}Al_{3}$ \cite{Degiorgi,Degiorgi2,Dressel}, have
shown no effects of the magnetic ordering on electronic properties at
frequencies above $30cm^{-1}$. However, recent infrared optical measurements 
\cite{Dressel2} show well pronounced (pseudo)gap behavior at rather low
frequency $\omega _{g}\approx 0.2$ $meV$. We stress that $\omega _{g}$ was
too small (compared to $T_{N}\approx 14$ $K$) in order to be explained with
the simple weak coupling SDW theory \cite{Grüner}, which predicts $2\omega
_{g}^{SDW}/T_{SDW}\approx 3.5$.

In connection with the above experimental results there is a necessity for a
thorough theoretical study of the optical properties of metals with the SDW
magnetic order. There is also an interesting question - how do impurities
and possible strong coupling effects affect the SDW state? This is the main
topic of this work. We stress also that in the following we will not discuss
a possible microscopic mechanism for the formation of the magnetic order.
This is a highly difficult problem which must additionally take into account
material specifics. Therefore, our analysis is based on a
semi-phenomenological model, which assumes that the conduction electrons are
affected by the magnetic order with some large wave-vector $Q(\sim 2k_{F})$.

In most part of the paper we deal with itinerant SDW magnetic systems, where
the SDW electron dynamics is governed by strong coupling equations which
also take into account impurity effects\ in strongly correlated systems -
Section II. A possible extension of the theory to the case of a localized
antiferromagnetic (LAF) order is discussed in Section II too. We point out
that heavy fermions belong to the class of strongly correlated electronic
systems in which charge fluctuations are strongly renormalized in such a way
that the forward scattering peak appears in scattering potentials \cite
{KuZe,Zey,Kulic}. This fact is taken into account in this work by
considering impurity effects on the SDW state. It turns out that in such a
case the renormalized impurities act as ''SDW-breaking'', and the latter
fact might play an important role in explanation of the optical properties
of SDW (or LAF) heavy fermions. We emphasize that the consideration of
strong coupling effects in some respect resembles the Eliashberg theory for
superconductivity. The single-particle dynamical conductivity of the SDW
state is calculated in Section III by including impurity and strong coupling
effects, while in Section IV the numerical calculations for $N(\omega )$, $%
\sigma (\omega )$ and $R(\omega )$are presented and discussed.

\section{Model of the SDW metal}

\subsection{Weak-coupling limit}

In the following we study the effect of the SDW magnetic order characterized
by the wave vector ${\bf Q}$ and by the effective exchange field $h_{\sigma ,%
{\bf Q}}=\sigma h$ which acts on the electronic spins. Since we study here
problems which are related to charge fluctuations in the framework of single
particle excitations, it turns out that nothing depends explicitly on the
spin index $\sigma $, i.e. there is simply a summation over $\sigma $. The
SDW\ Hamiltonian in the weak coupling limit is given by 
\begin{equation}
\hat{H}_{SDW}=\sum_{{\bf k}\sigma }[\xi _{{\bf k}}f_{{\bf k}\sigma
}^{\dagger }f_{{\bf k}\sigma }+h_{{\bf Q}\sigma }f_{{\bf k+Q}\sigma
}^{\dagger }f_{{\bf k}\sigma }]+c.c.,  \label{h}
\end{equation}
where $h_{{\bf Q}\sigma }=\sigma h_{{\bf Q}}$, and $\sigma =\pm $ . The bare
electronic excitation spectrum $\xi _{{\bf k}}=\epsilon _{{\bf k}}-\mu $ is
spin-independent. It is easy to diagonalize this Hamiltonian by the
Bogoliubov transformation $a_{{\bf k}\sigma 1}=u_{{\bf k}}f_{{\bf k}\sigma
}-\sigma v_{{\bf k}}f_{{\bf k+Q}\sigma }$ and $a_{{\bf k}\sigma 2}=\sigma v_{%
{\bf k}}f_{{\bf k}\sigma }+u_{k}f_{{\bf k+Q}\sigma }$. Here, $u_{{\bf k}%
}^{2}+v_{{\bf k}}^{2}=1$ and $u_{{\bf k}}^{2}=[1+\rho _{{\bf k}}/\sqrt{\rho
_{{\bf k}}^{2}+h_{{\bf Q}}^{2}}]/2$, where $\varepsilon _{{\bf k}}=(\xi _{%
{\bf k}}+\xi _{{\bf k+Q}})/2$ and $\rho _{{\bf k}}=(\xi _{{\bf k}}-\xi _{%
{\bf k+Q}})/2$ which gives the SDW quasiparticle spectrum $E_{{\bf k}%
,1/2}=\varepsilon _{{\bf k}}\mp \sqrt{\rho _{{\bf k}}^{2}+h_{{\bf Q}}^{2}}$.
Below we will study the case with appreciably nested spectrum 
\begin{equation}
\xi _{{\bf k}}\approx -\xi _{{\bf k+Q}}  \label{nesting}
\end{equation}
which gives the quasiparticle spectrum 
\begin{equation}
E_{{\bf k},n}=\mp E_{{\bf k}}=\mp \sqrt{\xi _{{\bf k}}^{2}+h_{{\bf Q}}^{2}}.
\label{spec}
\end{equation}

In that case the physics of the SDW system resembles the physics of
superconductivity. The experience with the latter implies that the SDW
coherence effects (the dependence on $u_{{\bf k}}$, $v_{{\bf k}}$) play
important role in various transport properties. Since the nonmagnetic
(optical) processes in SDW systems will be studied, we exploit this analogy
by introducing the Nambu (''spinor'') operator \ 
\begin{equation}
\Psi _{{\bf k}\sigma }^{\dagger }=(f_{{\bf k}\sigma }^{\dagger }f_{{\bf k+Q}%
\sigma }^{\dagger })  \label{nambu}
\end{equation}
and analogously the (column) operator $\Psi _{{\bf k}}$. In this notation
SDW Hamiltonian has the superconductivity-like form 
\begin{equation}
\hat{H}_{SDW}=\sum_{{\bf k}\sigma }\Psi _{{\bf k}\sigma }^{\dagger }[\xi _{%
{\bf k}}\hat{\tau}_{3}+\Delta ^{SDW}\hat{\tau}_{1}]\Psi _{{\bf k}\sigma },
\label{hnambu}
\end{equation}
where $\hat{\tau}_{i}$, $i=0,1,2,3$, are Pauli matrices ($\hat{\tau}_{0}=%
\hat{1}$) in Nambu space. In order to pursue the superconductivity analogy
we put $h_{{\bf Q\sigma }}\equiv \Delta ^{SDW}$ and assume it to be real. In
that case the Gor'kov equations for the Fourier transform $\hat{G}^{0}({\bf k%
},\omega _{n})$ of $\hat{G}_{\sigma \sigma }^{0}({\bf k},\tau )=-<T_{\tau
}\Psi _{{\bf k}\sigma }(\tau )\Psi _{{\bf k}\sigma }^{\dagger }(0)>$ ($\tau $
is the imaginary time) fulfills the standard weak-coupling BCS-like equation
(we omit the spin index $\sigma $)

\begin{equation}
\lbrack i\omega _{n}\hat{\tau}_{0}-\xi _{{\bf k}}\hat{\tau}_{3}-\Delta ^{SDW}%
\hat{\tau}_{1}]\hat{G}^{0}({\bf k},\omega _{n})=1  \label{gorkov}
\end{equation}
which has the solution 
\begin{equation}
\hat{G}^{0}({\bf k},\omega _{n})=-\frac{i\omega _{n}\hat{\tau}_{0}+\xi _{%
{\bf k}}\hat{\tau}_{3}+\Delta ^{SDW}\hat{\tau}_{1}}{\omega _{n}^{2}+E_{{\bf k%
}}^{2}},  \label{green}
\end{equation}
where $E_{{\bf k}}=\sqrt{\xi _{{\bf k}}^{2}+(\Delta ^{SDW})^{2}}$. If one
assumes that the magnetic order is of the SDW type then $\Delta \equiv h_{%
{\bf Q}}$ fulfills the BCS like self-consistent equation 
\begin{equation}
\Delta ^{SDW}=V_{SDW}\sum_{{\bf k},\omega _{n}}G_{12}^{0}({\bf k},\omega
_{n}).  \label{s-c}
\end{equation}
Below we omit the index SDW in $\Delta ^{SDW}$.

\subsection{Strong-coupling generalization and impurity effects}

(a) {\it Strong coupling effects}

The dynamics of the SDW order parameter can be generalized in order to
include possible strong coupling (retardation) effects, which might be due
to quasiparticle interaction with some bosonic (collective) excitations. The
generalization to the strong coupling SDW dynamics is analogous to the
Eliashberg theory for superconductors, where $\omega _{n}\rightarrow \tilde{%
\omega}_{n{\bf k}}$, $\Delta ^{SDW}\rightarrow \tilde{\Delta}^{SDW}(\omega
_{n},{\bf k})\equiv \tilde{\Delta}_{n{\bf k}}$ while the microscopic origin
of the SDW is contained in the interaction term $V_{SDW}\rightarrow
V_{SDW}(\omega _{n},{\bf k})$. One should underline that since the SDW order
is due to the Coulomb interaction vertex corrections to the Eliashberg
equations might be important. In the strong-coupling case $\hat{G}({\bf k}%
,\omega _{n})$ reads (electron-hole symmetry is assumed and the index SDW is
omitted) 
\[
\hat{G}({\bf k},\omega _{n})=-\frac{i\tilde{\omega}_{n{\bf k}}\hat{\tau}%
_{0}+\xi _{{\bf k}}\hat{\tau}_{3}+\tilde{\Delta}_{n{\bf k}}\hat{\tau}_{1}}{%
\tilde{\omega}_{n{\bf k}}^{2}+\xi _{{\bf k}}^{2}+\tilde{\Delta}_{n{\bf k}%
}^{2}},
\]
where $\tilde{\omega}_{n{\bf k}}$ and $\tilde{\Delta}_{n{\bf k}}$ are
solutions of the SDW-Eliashberg equations (we neglect a shift of the
chemical potential) 
\begin{equation}
\tilde{\omega}_{n{\bf k}}=\omega _{n}+\pi T\sum_{m,{\bf q}}V_{Z}(\omega
_{n}-\omega _{m},{\bf k}-{\bf q})\frac{\tilde{\omega}_{n{\bf q}}}{\tilde{%
\omega}_{n{\bf q}}^{2}+\xi _{{\bf q}}^{2}+\tilde{\Delta}_{n{\bf q}}^{2}},
\label{z}
\end{equation}
\begin{equation}
\tilde{\Delta}_{n{\bf k}}=\pi T\sum_{m{\bf q}}V_{SDW}(\omega _{n}-\omega
_{m},{\bf k}-{\bf q})\frac{\tilde{\Delta}_{n{\bf q}}}{\tilde{\omega}_{n{\bf q%
}}^{2}+\xi _{{\bf q}}^{2}+\tilde{\Delta}_{n{\bf q}}^{2}}.  \label{d}
\end{equation}
Here, the quasiparticle renormalization is described by all scattering
processes, i.e. $V_{Z}(\omega _{n},{\bf k})=V_{EP}(\omega _{n},{\bf k}%
)+V_{C}(\omega _{n},{\bf k})+...$ (EP - the electron-phonon interaction, C -
the Coulomb interaction, etc. ), while the $V_{SDW}(\omega _{n},{\bf k})$
interaction is responsible for the formation of the SDW condensate.

(b) {\it Effect of impurities}

The effect of nonmagnetic impurities is described by the Hamiltonian 
\begin{equation}
\hat{H}_{imp}=\sum_{i,{\bf k}}\rho _{i}({\bf k})V_{i}({\bf k})\hat{\rho}(%
{\bf k})  \label{imp}
\end{equation}
with 
\begin{equation}
\hat{\rho}({\bf k})=\sum_{{\bf q}\sigma }f_{{\bf q+k}\sigma }^{\dagger }f_{%
{\bf q}\sigma }.  \label{ro}
\end{equation}
In strongly correlated systems like heavy fermions the backward scattering
in the charge scattering potential is strongly suppressed \cite
{KuZe,Zey,Kulic}, i.e. $V_{i}({\bf k})$ is appreciable for $\mid {\bf k\mid }
$ $<k_{c}\ll k_{F},Q$. We simplify the calculations by assuming $V_{i}({\bf k%
})=V_{io}=const$ for $\mid {\bf k\mid }$ $<k_{c}$. In that case the leading
term in $\hat{H}_{imp}$ is the SDW-breaking part, i.e. $\hat{H}_{imp}=\hat{H}%
_{imp}^{SDW-b}+$... 
\begin{equation}
\hat{H}_{imp}^{SDW-b}=\sum_{i,{\bf k}}\rho _{i}({\bf k})V_{i0}\sum_{{\bf q}%
\sigma }\Psi _{{\bf q+k}\sigma }^{\dagger }\hat{\tau}_{0}\Psi _{{\bf q}%
\sigma }.  \label{SDW-b}
\end{equation}
In the presence of such impurities the SDW-Eliashberg equations read

\[
\tilde{\omega}_{n{\bf k}}=\omega _{n}+n_{i}V_{0}^{2}\sum_{{\bf q}}\frac{%
\tilde{\omega}_{n{\bf q}}}{\tilde{\omega}_{n}^{2}+\xi _{{\bf q}}^{2}+\tilde{%
\Delta}_{n{\bf q}}^{2}}+
\]
\begin{equation}
+\pi T\sum_{m{\bf q}}V_{Z}(\omega _{n}-\omega _{m},{\bf k}-{\bf q})\frac{%
\tilde{\omega}_{m{\bf q}}}{\tilde{\omega}_{m{\bf q}}^{2}+\xi _{{\bf q}}^{2}+%
\tilde{\Delta}_{m{\bf q}}^{2}},  \label{om}
\end{equation}
\multe
\begin{equation}
\tilde{\Delta}_{n{\bf k}}=-n_{i}V_{0}^{2}\sum_{{\bf q}}\frac{\tilde{\Delta}%
_{n{\bf q}}}{\tilde{\omega}_{n{\bf q}}^{2}+\xi _{{\bf q}}^{2}+\tilde{\Delta}%
_{n{\bf q}}^{2}}+\pi T\sum_{m{\bf q}}V_{SDW}(\omega _{n}-\omega _{m},{\bf k}-%
{\bf q})\frac{\tilde{\Delta}_{m{\bf q}}}{\tilde{\omega}_{m{\bf q}}^{2}+\xi _{%
{\bf q}}^{2}+\tilde{\Delta}_{m{\bf q}}^{2}}.  \label{di}
\end{equation}
\multb
The above equations are suitable for analyzing very anisotropic SDW order
parameters, even unconventional ones. Some properties of uncoventional SDW
(USDW) in the weak-coupling and clean ($n_{i}=0$) limit were studied in \cite
{Dora,Xiao}.

In the following a small anisotropy of the spectral functions $V_{Z}(\omega
_{m},{\bf k})$ and $V_{SDW}(\omega ,{\bf k})$ is assumed. After the
integration of the above equations over energy one obtains 
\multe
\begin{equation}
\tilde{\omega}_{n}=\omega _{n}+\gamma _{imp}\frac{\tilde{\omega}_{n}}{\sqrt{%
\tilde{\omega}_{n}^{2}+\tilde{\Delta}_{n}^{2}}}+\pi T\sum_{m}\lambda
_{Z}(\omega _{n}-\omega _{m})\frac{\tilde{\omega}_{n}}{\sqrt{\tilde{\omega}%
_{n}^{2}+\tilde{\Delta}_{n}^{2}}},  \label{om1}
\end{equation}

\begin{equation}
\tilde{\Delta}_{n}=-\gamma _{imp}\frac{\tilde{\Delta}_{n}}{\sqrt{\tilde{%
\omega}_{n}^{2}+\tilde{\Delta}_{n}^{2}}}+\pi T\sum_{m}\lambda _{SDW}(\omega
_{n}-\omega _{m})\frac{\tilde{\Delta}_{n}}{\sqrt{\tilde{\omega}_{n}^{2}+%
\tilde{\Delta}_{n}^{2}}},  \label{di2}
\end{equation}
\multb
where $\gamma _{imp}(=2\pi n_{i}N(0)V_{0}^{2})$ is\ the impurity scattering
rate. Note, that $\gamma _{imp}$ enters equations for $\tilde{\omega}_{n{\bf %
k}}$ and $\tilde{\Delta}_{n{\bf k}}$ with different signs, due to the
detrimental (breaking) effect of nonmagnetic impurities on the SDW state.
The coupling functions $\lambda _{l}(\omega _{n}-\omega _{m})=\lambda
_{\{Z,SDW\}}(\omega _{n}-\omega _{m})$ \ depend on the spectral functions $%
\alpha _{l}^{2}F_{l}(\omega )$ 
\begin{equation}
\lambda _{l}(\omega _{n}-\omega _{m})=\int_{0}^{\infty }d\omega \alpha
_{l}^{2}F_{l}(\omega )\frac{2\omega }{\omega ^{2}+(\omega _{n}-\omega
_{m})^{2}}.  \label{lambda}
\end{equation}
In our numerical calculations we assume $\lambda _{Z}=$ $\lambda _{SDW}$
only because of simplicity.

We would like to point out that the above theory for the SDW systems can be
extended to electronic systems in the presence of a localized AF magnetic
(LAF) order which is certainly governed by the microscopic dynamics which is
different from the SDW one. In the case of the LAF magnetic order the
self-consistent equation Eq.(\ref{di2}) does not hold. However, \ the
nonmagnetic impurities affect the quasiparticle Green's function by changing
effectively the LAF order parameter in the following way 
\begin{equation}
\tilde{\Delta}_{n}=\Delta _{0}(T)-\gamma _{imp}\frac{\tilde{\Delta}_{n}}{%
\sqrt{\tilde{\omega}_{n}^{2}+\tilde{\Delta}_{n}^{2}}},  \label{diAF}
\end{equation}
where the temperature dependent LAF order parameter $\Delta _{0}(T)(\equiv
h_{{\bf Q}})$ is the solution of some unknown self-consistency equation. The
latter equation should take into account the spin dynamics of real
materials. We stress that for numerical calculations of $N(\omega )$, $%
\sigma (\omega )$ and $R(\omega )$ the temperature dependence of $\Delta
_{0}(T)$ can be extracted from magnetic neutron scattering experiments.

\section{Dynamical conductivity of the SDW metal}

We calculate the (diagonal) dynamical conductivity $\sigma _{jj}(\omega )$ $%
(j=x,y,z)$, which is given by 
\begin{equation}
\sigma _{jj}(\omega )=\frac{i\Pi _{jj}({\bf q}\rightarrow 0,i\omega
_{n}\Longrightarrow \omega +i0^{+})}{\omega }  \label{sig}
\end{equation}
where $\Pi (i\omega _{n})$ is the current-current correlation function 
\begin{equation}
\Pi _{jj}({\bf q},i\omega _{n})=\int_{0}^{\beta }d\tau e^{i\omega _{n}\tau
}<T\hat{\jmath}_{j}({\bf q},\tau )\hat{\jmath}_{j}(-{\bf q,}0)>.  \label{pi}
\end{equation}
Since we study the optical properties in the London limit (${\bf %
q\rightarrow 0}$) then in that limit the current operator $\hat{\jmath}_{j}(%
{\bf q},\tau )$ of the SDW metal reads 
\begin{equation}
\hat{\jmath}_{j}({\bf q},\tau )=\frac{e}{m}\sum_{\sigma ,{\bf p}}V_{j}({\bf %
p+}\frac{{\bf q}}{2})\Psi _{{\bf p}\sigma }^{\dagger }\hat{\tau}_{3}\Psi _{%
{\bf p+q}\sigma },  \label{j}
\end{equation}
where in obtaining $\hat{\jmath}_{j}({\bf q},\tau )$ the property ${\bf V}%
_{F}({\bf k}_{F})=-{\bf V}_{F}({\bf k}_{F}+{\bf Q})$ of the Fermi velocity
was used. We point out that there is a significant difference between the
expressions for the current (for ${\bf q\rightarrow 0}$) in the SDW state -
it contains $\hat{\tau}_{3}$, and that in the superconducting (SC) state -
it contains $\hat{\tau}_{0}$. This fact causes profound differences in
transport properties of these two systems. In a separate work \cite{Oleg1}
it will be shown that the SDW dynamical conductivity ($\sigma _{SDW}({\bf %
q\rightarrow 0,}\omega ,T)$) behaves like the phonon self-energy (up to some
sign factors) in superconductors, i.e. $\sigma _{SDW}(\omega ,T)\sim \Sigma
_{SC}^{ph}(\omega ,T)$ and vice versa $\Sigma _{SDW}^{ph}(\omega ,T)\sim
\sigma _{SC}(\omega ,T)$.

By neglecting the vertex correction in $\Pi _{jj}({\bf q},i\omega _{n})$ one
obtains 
\begin{equation}
\Pi _{jj}^{SDW}({\bf q=0},i\omega _{n})\approx \frac{\omega _{pl}^{2}}{4\pi }%
T\sum_{m\sigma }\Pi _{jj\sigma }^{SDW}(\omega _{n},\omega _{m}),
\label{pi-app}
\end{equation}
\multe
\begin{equation}
\Pi _{jj\sigma }^{SDW}(\omega _{n},\omega _{m})=\int d\xi _{{\bf k}}Sp\{\hat{%
\tau}_{3}\hat{G}_{\sigma \sigma }({\bf k},\omega _{n}+\omega _{m})\hat{\tau}%
_{3}\hat{G}_{\sigma \sigma }({\bf k},\omega _{m})\}.  \label{pi-nm}
\end{equation}
\multb
(The vertex correction to $\Pi _{jj}^{SDW}$ gives rise to the transport
scattering rate $\gamma _{tr}(\omega )$, i.e. to the renormalization $\gamma
(\omega )\rightarrow \gamma _{tr}(\omega ))$. After the integration over the
energy variable one obtains

\begin{equation}
\Pi _{jj}^{SDW}(\omega _{n},\omega _{m})=\frac{\tilde{\omega}_{m}(\tilde{%
\omega}_{m}+\tilde{\omega}_{m+n})+\tilde{\Delta}_{m}(\tilde{\Delta}_{m}+%
\tilde{\Delta}_{m+n})}{R_{m}P_{mn}}-  \label{pi2-nm}
\end{equation}
\[
-\frac{\tilde{\omega}_{m+n}(\tilde{\omega}_{m+n}+\tilde{\omega}_{m})+\tilde{%
\Delta}_{m+n}(\tilde{\Delta}_{m+n}+\tilde{\Delta}_{m})}{R_{m}P_{mn}}\text{ } 
\]
with $R_{m}=\sqrt{\tilde{\omega}_{m}^{2}+\tilde{\Delta}_{m}^{2}}$ and $%
P_{mn}=\tilde{\omega}_{m}^{2}-\tilde{\omega}_{m+n}^{2}+\tilde{\Delta}%
_{m}^{2}-\tilde{\Delta}_{m+n}^{2}$. If one compares this expression with the
corresponding one for superconductors\cite{Bickers} one sees that $\Pi
_{jj}^{SDW}$ contains the coherence factors $\tilde{\Delta}_{m}(\tilde{\Delta%
}_{m}+\tilde{\Delta}_{m+n})$ and $\tilde{\Delta}_{m+n}(\tilde{\Delta}_{m+n}+%
\tilde{\Delta}_{m})$ while in the SC state one has $\tilde{\Delta}_{m}(%
\tilde{\Delta}_{m}-\tilde{\Delta}_{m+n})$ and $\tilde{\Delta}_{m+n}(\tilde{%
\Delta}_{m+n}-\tilde{\Delta}_{m})$.

In order to obtain the optical conductivity we have to make an analytical
continuation of Eq.(\ref{pi2-nm}) to the real frequency axis 
\multe
\[
\sigma _{jj}^{SDW}(\omega )=\frac{\omega _{j,pl}^{2}}{4\pi \omega }\left.
\left\{ i\pi T\sum_{n}\Pi _{jj}^{SDW}(\omega _{n},\omega _{m})\right\}
\right| _{i\omega _{m}\Longrightarrow \omega +i0^{+}},
\]
where $\omega _{j,pl}=(8\pi e^{2}N(0)\left\langle V_{j}^{2}\right\rangle
)^{1/2}$ is the bare plasma frequency. The analytical continuation gives 
\begin{eqnarray}
\sigma _{jj}^{SDW}(\omega ) &=&\frac{i\omega _{j,pl}^{2}}{16\pi \omega }\int
d\omega ^{\prime }\left\{ \frac{\tanh \left( \frac{\omega _{-}}{2T}\right) }{%
D^{R}}\right. \left[ 1-\frac{\tilde{\omega}_{-}^{R}\tilde{\omega}_{+}^{R}%
{\bf -}\tilde{\Delta}_{-}^{R}\tilde{\Delta}_{+}^{R}}{\sqrt{(\tilde{\omega}%
_{\_}^{R})^{2}{\bf -(}\tilde{\Delta}_{-}^{R})^{2}}\sqrt{(\tilde{\omega}%
_{+}^{R})^{2}{\bf -(}\tilde{\Delta}_{+}^{R})^{2}}}\right] -  \label{real} \\
&&\frac{\tanh \left( \frac{\omega _{+}}{2T}\right) }{D^{A}}\left[ 1-\frac{%
\tilde{\omega}_{-}^{A}\tilde{\omega}_{+}^{A}{\bf -}\tilde{\Delta}_{-}^{A}%
\tilde{\Delta}_{+}^{A}}{\sqrt{(\tilde{\omega}_{\_}^{A})^{2}{\bf -(}\tilde{%
\Delta}_{-}^{A})^{2}}\sqrt{(\tilde{\omega}_{+}^{A})^{2}{\bf -(}\tilde{\Delta}%
_{+}^{A})^{2}}}\right] -  \nonumber \\
&&\frac{\tanh \left( \frac{\omega _{+}}{2T}\right) -\tanh \left( \frac{%
\omega _{-}}{2T}\right) }{D^{a}}\left. \left[ 1-\frac{\tilde{\omega}_{-}^{A}%
\tilde{\omega}_{+}^{R}{\bf -}\tilde{\Delta}_{-}^{A}\tilde{\Delta}_{+}^{R}}{%
\sqrt{(\tilde{\omega}_{\_}^{A})^{2}{\bf -(}\tilde{\Delta}_{-}^{A})^{2}}\sqrt{%
(\tilde{\omega}^{R})^{2}{\bf -(}\tilde{\Delta}_{+}^{R})^{2}}}\right]
\right\} ,  \nonumber
\end{eqnarray}
\multb
where 
\[
D^{R,A}=\sqrt{(\tilde{\omega}_{+}^{R,A})^{2}{\bf -(}\tilde{\Delta}%
_{+}^{R,A})^{2}}+\sqrt{(\tilde{\omega}_{\_}^{R,A})^{2}{\bf -(}\tilde{\Delta}%
_{-}^{R,A})^{2}},
\]
and 
\[
D^{a}=\sqrt{(\tilde{\omega}_{+}^{R})^{2}{\bf -(}\tilde{\Delta}_{+}^{R})^{2}}-%
\sqrt{(\tilde{\omega}_{\_}^{A})^{2}{\bf -(}\tilde{\Delta}_{-}^{A})^{2}}.
\]
$\omega _{\pm }$ denotes $\omega ^{\prime }\pm \omega /2$, and indices $R(A)$
correspond to the retarded (advanced) branches of the complex functions $%
F^{R(A)}=%
\mathop{\rm Re}%
F\pm i%
\mathop{\rm Im}%
F$. These expressions are similar to the ones for the optical conductivity
of strong-coupling superconductors (see\cite{Nam}) but with different sign
in the term $\tilde{\Delta}_{-}^{R,A}\tilde{\Delta}_{+}^{R,A}$ in the
coherence factors. This fact was used to describe experimental data in
systems with C(S)DW by using the famous Mattis-Bardeen expressions for the
optical conductivity with the coherence factor of the first kind\cite{DD}.
It is well known that these formulas work properly in two limits: a) in the
nonlocal Pippard one and b) in the very dirty case. But none of these limits
is valid for SDW in heavy-fermion systems because they are in the
normal-skin (local) limit, and as mentioned \ before normal impurities are
pair-breaking and destroy SDW-itself.

We note that $\sigma _{jj}^{SDW}(\omega )$ is derived by assuming that the
whole Fermi surface is gapped in the SDW state, i.e. that below $T_{N}$ the
metal passes into the dielectric phase. However, in most of real heavy
fermions with the SDW (or LMF) magnetic order\ the optical measurements show
Drude-like behavior ($\sigma _{jj}^{D}$) at low frequencies $\omega \ll
2\Delta _{SDW}^{0}$, which shows that the Fermi surface is only partially
gapped. In that case it is qualitatively justified to assume that the total
conductivity is the sum of both components, $\sigma _{jj}\approx \sigma
_{jj}^{D}+\sigma _{jj}^{SDW}$. This problem will be discussed in the next
section.

\section{Numerical results and discussion}

In the following we calculate the tunneling density of states $N(\omega )$
and $\sigma ^{SDW}(\omega )$ for various scattering rates $\gamma _{imp}$ in
the SDW model with two types of spectral functions $\alpha
_{SDW}^{2}F_{SDW}(\omega )$ shown in Fig.1. Note, that $\alpha
_{SDW}^{2}F_{SDW}(\omega )$ in Fig.1 - the spectrum a) has a form
reminiscent of that used in spin-fluctuation theories of the high-$T_{c}$
physics (see reviews \cite{Kampf,Kulic}), while the spectral function b) $%
\alpha _{SDW}^{2}F_{SDW}(\omega )$ in Fig.1 mimics a slightly broadened
Einstein-like spectrum. These two spectral functions are chosen for
simplicity and it will be demonstrated below that the global structure of $%
\sigma ^{SDW}(\omega )$ is practically independent on their shapes.

(i) {\it Density of states }$N(\omega )$

The tunneling density of states $N(\omega )$ is presented in Fig.2 for the
a) spectrum $\alpha _{SDW}^{2}F_{SDW}(\omega )$ from Fig.1, for the coupling
constant $\lambda (\equiv \lambda ^{SDW}(\omega =0))=4;$ (see, Eq.(\ref
{lambda})). It is seen that in the clean limit ($\gamma _{imp}=0$) $N(\omega
)$ has a strong BCS-like singularity due to the gap in the quasiparticle
spectrum. As we already mentioned, in real SDW heavy fermion systems there
is no complete gapping of the Fermi surface due to incomplete nesting. In
the latter case there are also states inside the gap - we call this
pseudogap behavior. Since the (nonmagnetic) impurities with a forward
scattering peak in the scattering potential act as SDW-breaking they
strongly suppress $N(\omega )$, by smoothing it and shifting its maximum to
lower frequencies - Fig.2. For $\gamma _{imp}\sim \Delta _{0}$ the density
of states, $N(\omega )$, shows the tendency to be gapless - a property
analogous to the case of superconductors with magnetic impurities. In the
case of the b) spectral function from Fig.1 with a large coupling constant $%
\lambda =5$, there is a clear dip (with $N(\omega )/N_{n}(0)<1$) in $%
N(\omega )$ which is followed by a broad peak at larger frequency - Fig.3.
We emphasize that the dip structure, followed by the peak at higher
frequencies, appears always in the strong coupling Eliashberg theory of
superconductors for sufficiently large coupling constant $\lambda $ and it
is independent on the shape of the spectral function \cite{Scalapino}.
Similar behavior of the tunneling density of states $N(\omega )$ has been
seen also in the tunneling conductivity of $UPd_{2}Al_{3}$ in the
superconducting state \cite{Adrian}. Since the above equations for the Green
functions and $N(\omega )$ hold also for superconductors at $T<T_{c}<T_{N}$
the results shown in Fig.3 confirm earlier conclusions \cite{Adrian,Sato1},
that superconductivity in $UPd_{2}Al_{3}$ is in the very strong coupling
regime ($\lambda \gg 1$) and most probably due to the spin-fluctuations. It
would be interesting to search experimentally for possible strong coupling
effects (the dip and second peak) in SDW materials at temperatures below $%
T_{N}$ but above $T_{c}.$

Similar conclusions hold also for the properties of $N(\omega )$ in the case
of a) spectral function $\alpha _{SDW}^{2}F_{SDW}(\omega )$ - in Fig.1,
where strong coupling effects are also pronounced for $\lambda _{SDW}\gg 1$.

(ii) {\it Dynamical conductivity} $\sigma ^{SDW}(\omega )${\em \ }{\it and
reflectivity }$R(\omega )$

Due to the peculiar coherence effects in the SDW state the dynamical
conductivity $\sigma ^{SDW}(\omega )$ differs significantly from the
corresponding one in s-wave superconductors - see Eq.(\ref{real}).

The numerical calculations of $\sigma ^{SDW}(\omega )=%
\mathop{\rm Re}%
\sigma ^{SDW}(\omega )+i%
\mathop{\rm Im}%
\sigma ^{SDW}(\omega )$ are shown in Figs.(4-5) for the a) and b) spectral
functions $\alpha _{SDW}^{2}F_{SDW}(\omega )$ in Fig.1. In the clean limit ($%
\gamma _{imp}=0$) and at very low temperatures ($T\ll \Delta _{0}$) $%
\mathop{\rm Re}%
\sigma ^{SDW}(\omega )$ (and $%
\mathop{\rm Im}%
\sigma ^{SDW}(\omega )$) show a singularity at $\omega =2\Delta _{0}$, while
at finite $T$ $%
\mathop{\rm Re}%
\sigma ^{SDW}$ and $%
\mathop{\rm Im}%
\sigma ^{SDW}$ are broadened- see Fig.4$.$

In the case of a finite impurity scattering rate with $\gamma _{imp}<\Delta
_{0},$ both $%
\mathop{\rm Re}%
\sigma ^{SDW}$ and $%
\mathop{\rm Im}%
\sigma ^{SDW}$ are broadened and their maximum and minimum are shifted to
lower frequencies, as shown in Fig.5. It is apparent from Fig.5 that for $%
\gamma _{imp}=0.11\Delta _{0}$ the suppression, broadening and shift (to
lower frequencies) of $\sigma ^{SDW}$ are very pronounced, already for a
large mean-free path, $l\sim 20$ $\xi _{SDW}$, impurities affect the SDW
condensate appreciably. This property is robust and practically independent
on the shape of the spectral function. We stress that this pronounced
impurity dependence of $\sigma ^{SDW}(\omega )$ may serve as an important
check of the SDW state in heavy fermions.

The frequency dependence of the reflectivity, $R^{SDW}(\omega )$, in the SDW
state is shown in Fig.6 for the calculated conductivity $\sigma ^{SDW}$ in
Fig.5. Note, that the reflectivity along the j-axis $R_{j}(\omega )$ is
defined by the dielectric function $\varepsilon _{jj}(\omega )$

\begin{equation}
R_{j}(\omega )=\left| \frac{\sqrt{\varepsilon _{jj}(\omega )}-1}{\sqrt{%
\varepsilon _{jj}(\omega )}+1}\right| ^{2}  \label{R}
\end{equation}
\begin{equation}
\varepsilon _{jj}(\omega )=\varepsilon _{jj,\infty }+\frac{4\pi i\sigma
_{jj}^{SDW}(\omega )}{\omega }.  \label{eps}
\end{equation}
In the calculations it was assumed $\varepsilon _{jj,\infty }=9$. 

From Fig.6 a rich structure in $R(\omega )$ is seen which is due to the
strong coupling limit ($\lambda =5$) of the theory. In the clean case there
is a peak at $\omega =2\Delta _{0}$ followed by the minimum around $\omega
/\Omega \approx 5$ which is just the position of the dip in the density of
states shown in Fig.3, while the maximum of the shoulder at $\omega /\Omega
\approx 7.5$ corresponds to the second maximum in $N(\omega )$.

Formally one can model $\sigma ^{SDW}(\omega )$ with the generalized Drude
formula 
\begin{equation}
\sigma _{jj}^{SDW}(\omega )=\frac{\omega _{pl,jj}^{2}}{4\pi }\frac{1}{\Gamma
_{eff}(\omega )-i\omega m_{SDW}^{\ast }(\omega )/m_{\infty }}.
\label{DrudeSDW}
\end{equation}
In the case of SDW state one obtains that the effective (optical) mass, $%
m_{SDW}^{\ast }(\omega )$, is negative, while $\Gamma _{eff}(\omega )$ -
shown in Fig.7, has also a rich structure due to strong coupling effects
which are similar to the superconducting ones\cite{Comb}.

If one wants to compare the obtained results with experimental results on
SDW heavy fermions one should take into account that the Fermi surface in
real  heavy fermions is not fully gapped in the SDW (and LAF) states. This
means that part of the Fermi surface contributes to the Drude conductivity
for $\omega <2\Delta _{0}$. Such an analysis will depend on material
properties and will be studied elsewhere. However, in order to model this
situation we assume these two effects to be additive, i.e. $\sigma (\omega
)=\sigma ^{D}(\omega )+\sigma ^{SDW}(\omega )$ (we omit the index $j$) 
\begin{equation}
\sigma (\omega )=\frac{1}{4\pi }\frac{\omega _{pl,D}^{2}}{-i\omega +\gamma
_{D}}+\frac{1}{4\pi }\frac{\omega _{pl,SDW}^{2}}{-i\omega +W_{SDW}(\omega )},
\label{sig_tot1}
\end{equation}
where the optical self-energy $W_{SDW}(\omega )$ can be related to Eqs.(\ref
{real}, \ref{DrudeSDW}). Eq.(\ref{sig_tot1}) can be represented in the
generalized Drude form 
\begin{equation}
\sigma (\omega )=\frac{\omega _{pl,D}^{2}+\omega _{pl,SDW}^{2}}{4\pi }\frac{1%
}{-i\omega m^{\ast }(\omega )/m_{\infty }+\Gamma _{eff}(\omega )}.
\label{sig_tot2}
\end{equation}
The total complex conductivity $\sigma (\omega )$ is numerically calculated
by assuming $\gamma _{D}=2\gamma _{imp}$, $\omega _{pl,D}^{2}=\omega
_{pl,SDW}^{2}$ and is shown in Fig.8, while $m^{\ast }(\omega )/m_{\infty }$
and $\Gamma _{eff}(\omega )$ are shown in Fig.9. The rich structure in $%
N(\omega )$ and $\sigma (\omega )$ is reflected in $m^{\ast }(\omega
)/m_{\infty }$ and $\Gamma _{eff}(\omega )$. Even such a qualitative
analysis shows surprising similarity in structure of the above theoretical
predictions and experimental results in $UPd_{2}Al_{3}$ \cite{Dressel2}. A
similar conclusion holds also for the experimental and theoretical total
reflectivity $R(\omega )$, where the latter is shown in Fig.10. We note
again, that in order to compare to experimental results in real heavy
fermion materials one should know realistic values of $\omega _{pl,D}^{2}$, $%
\omega _{pl,SDW}^{2}$, $m_{\infty }$ etc. This will be analyzed elsewhere. 

In conclusion, we have studied strong coupling and impurity effects on the
dynamical conductivity $\sigma (\omega )\ $and tunneling density of states $%
N(\omega )$\ in heavy fermion metals with SDW magnetic order. We have shown
that nonmagnetic impurities, which have pronounced forward scattering peak
in the impurity potential in strongly correlated systems - like heavy
fermions, strongly affect $N(\omega )$, $\sigma (\omega )$ and $R(\omega )$.
Such a SDW-breaking behavior might be a relevant test for the itinerant
character of SDW magnetic order and the forward impurity scattering peak in
the nonmagnetic impurity potential of heavy fermions.

\acknowledgments  The authors are thankful to B. Gorshunov for fruitful
discussions. M. L. K. acknowledges the support of the Deutsche
Forschungsgemeinschaft through the SFB 464 project at University of Augsburg.

\begin{figure}[tbp]
\caption{Spectral function $\protect\alpha_{SDW}^{2}F_{SDW}(\protect\omega)$%
: a) spin-fluctuation spectrum, b) Einstein-like spectrum. Peaks are at $%
\protect\omega=\Omega$.}
\label{fig1}
\epsfig{file=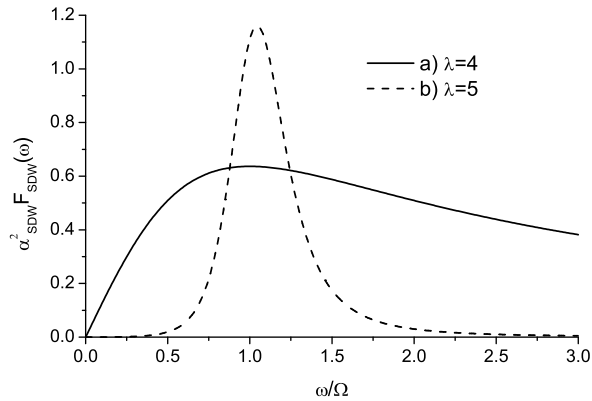,width=\linewidth,clip=true}
\end{figure}
\begin{figure}[tbp]
\caption{Tunneling density of states $N(\protect\omega)$ for the spectrum in
Fig.1a with $\protect\lambda^{SDW}=4$ for various impurity scattering rates $%
\protect\gamma_{imp}$.}
\label{fig2}
\epsfig{file=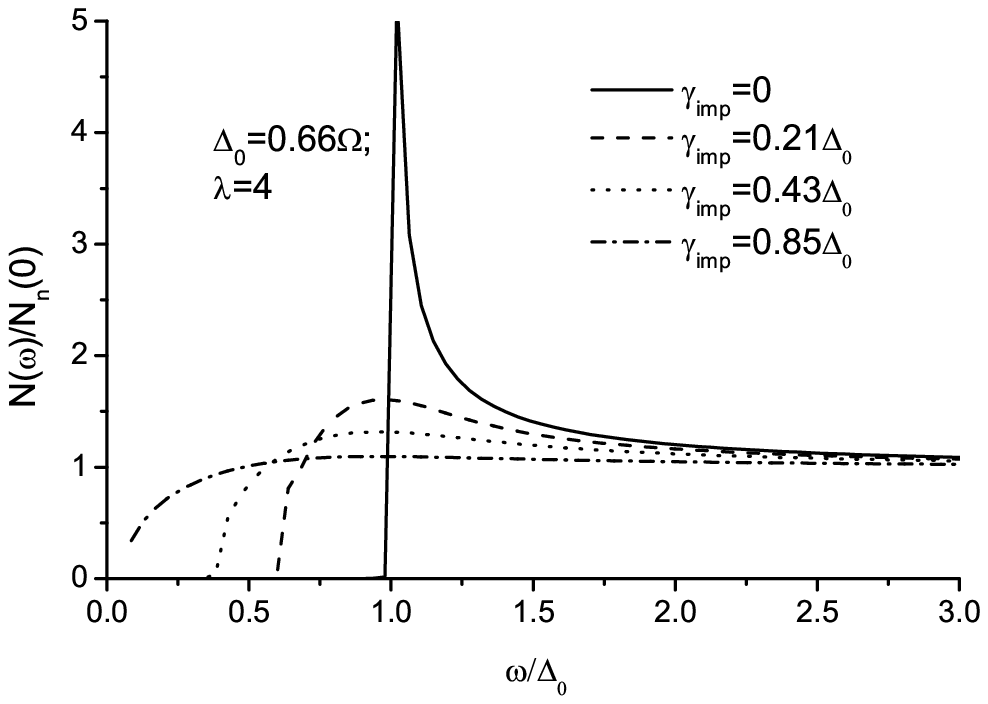,width=\linewidth,clip=true}
\end{figure}
\begin{figure}[tbp]
\caption{Tunneling density of states $N(\protect\omega)$ for the spectrum in
Fig.1b with $\protect\lambda^{SDW}=5$ for various impurity scattering rates $%
\protect\gamma_{imp}$.}
\label{fig3}
\epsfig{file=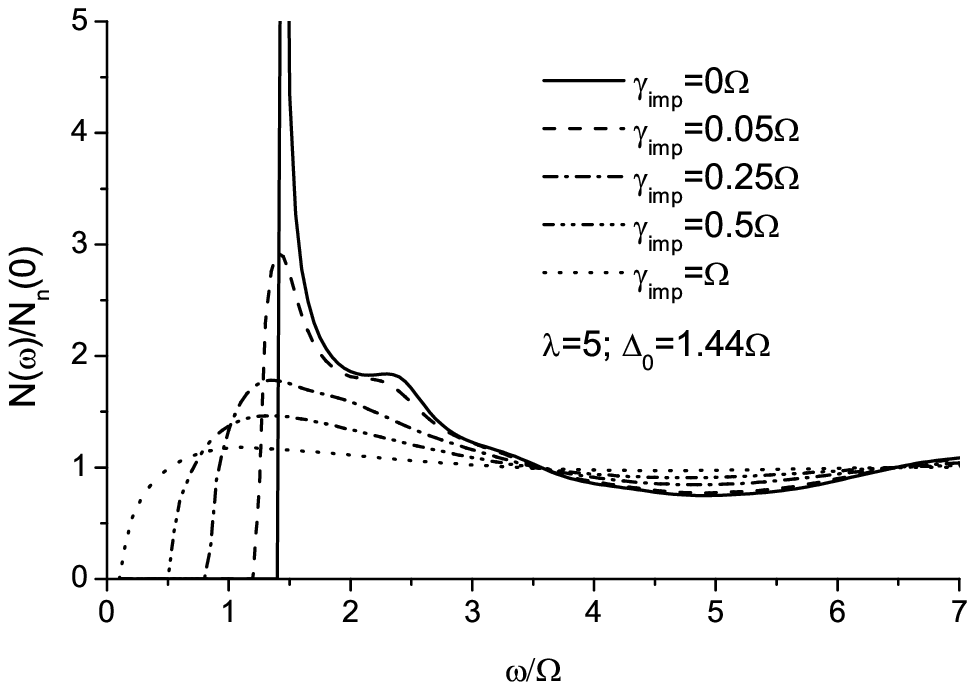,width=\linewidth,clip=true}
\end{figure}
\newpage
\begin{figure}[tbp]
\caption{Complex dynamical conductivity $\protect\sigma^{SDW}(\protect\omega)
$ for the spectrum in Fig.1a with $\protect\lambda^{SDW}=4$ for various
temperatures $T$ in the clean limit $\protect\gamma_{imp}=0$.}
\label{fig4}
\epsfig{file=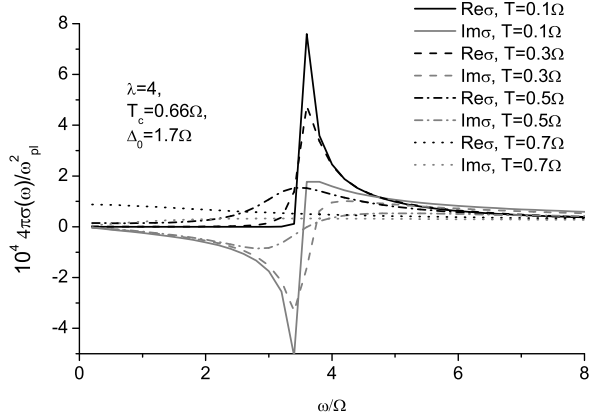,width=\linewidth,clip=true}
\end{figure}
\begin{figure}[tbp]
\caption{Complex dynamical conductivity $\protect\sigma^{SDW}(\protect\omega)
$ for the spectrum in Fig.1b with $\protect\lambda ^{SDW}=5$ for various
impurity scattering rates $\protect\gamma_{imp}$.}
\label{fig5}
\epsfig{file=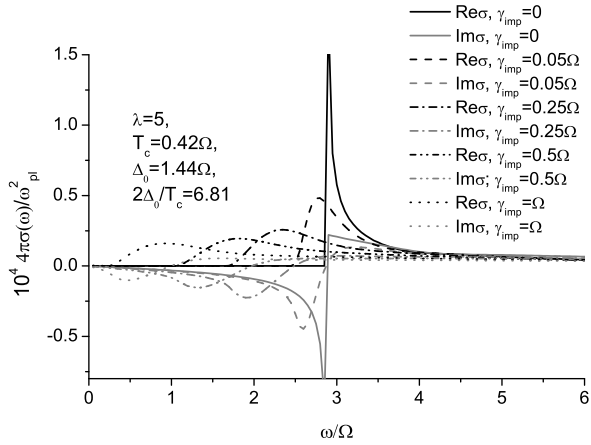,width=\linewidth,clip=true}
\end{figure}
\begin{figure}[tbp]
\caption{Reflectivity $R^{SDW}(\protect\omega)$ for $\protect\sigma^{SDW}(%
\protect\omega)$ in Fig.5.}
\label{fig6}
\epsfig{file=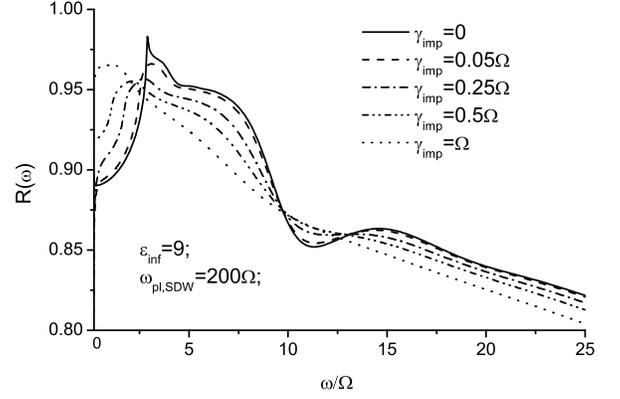,width=\linewidth,clip=true}
\end{figure}
\begin{figure}[tbp]
\caption{Optical relaxation rate $\Gamma_{eff}^{SDW}(\protect\omega )$ - see
Eq.(\ref{DrudeSDW}) for spectral function in Fig.1b with $\protect\lambda =5$
and for various impurity scattering rates $\protect\gamma _{imp}$.}
\label{fig7}
\epsfig{file=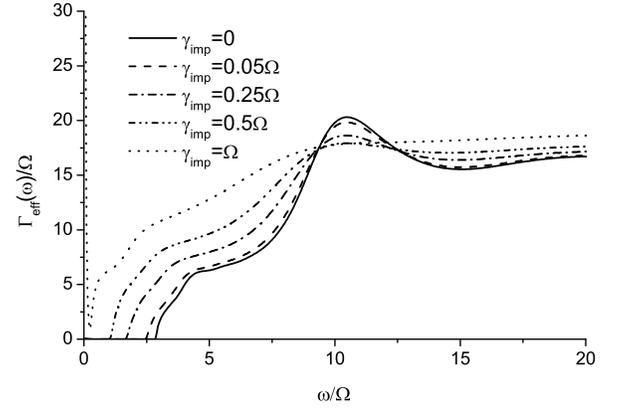,width=\linewidth,clip=true}
\end{figure}
\begin{figure}[tbp]
\caption{Total conductivity $\protect\sigma(\protect\omega)=\protect\sigma%
^{D}(\protect\omega)+\protect\sigma ^{SDW}(\protect\omega)$ for $\protect%
\sigma^{SDW}(\protect\omega)$ in Fig.5 and for various impurity scattering
rates $\protect\gamma _{imp}$.}
\label{fig8}
\epsfig{file=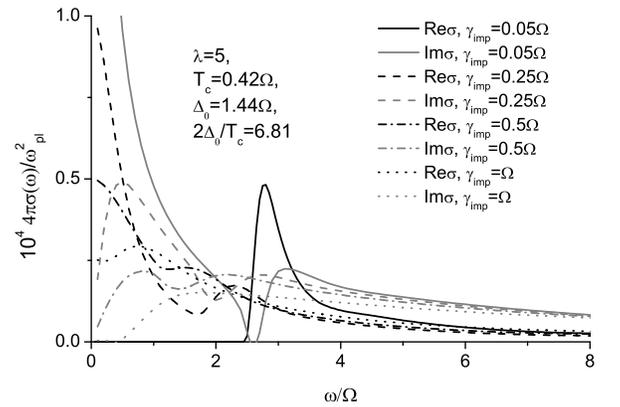,width=\linewidth,clip=true}
\end{figure}
\begin{figure}[tbp]
\caption{Optical relaxation rate $\Gamma_{eff}(\protect\omega)$ and the
effective mass $m^{\ast}(\protect\omega)$ from total reflectivity in Fig.8 -
see Eq.(\ref{sig_tot2}) for spectral function in Fig.1b with $\protect%
\lambda =5$ and for various impurity scattering rates $\protect\gamma_{imp}$%
. }
\label{fig9}
\epsfig{file=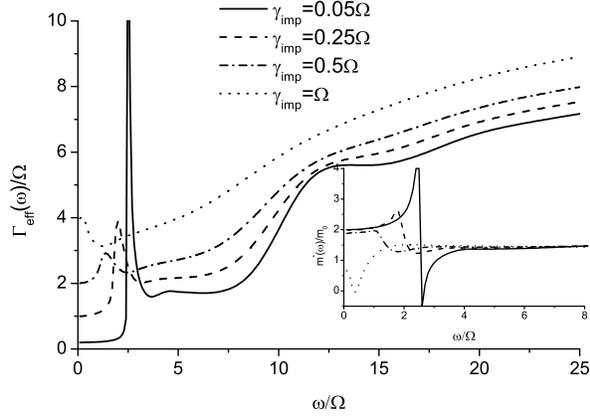,width=\linewidth,clip=true}
\end{figure}
\begin{figure}[tbp]
\caption{Total reflectivity $R(\protect\omega)=R^{D}(\protect\omega)+R^{SDW}(%
\protect\omega)$ for $\protect\sigma^{SDW}(\protect\omega)$ in Fig.5 and for
various impurity scattering rates $\protect\gamma _{imp}$.}
\label{fig10}
\epsfig{file=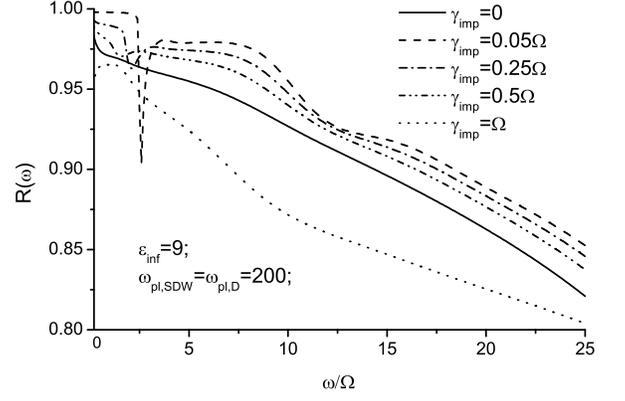,width=\linewidth,clip=true}
\end{figure}
\multe

\end{document}